\documentclass[12pt,preprint]{aastex}

\shorttitle{Properties of solar ephemeral regions}
\shortauthors{Yang et al.}

\begin{document}

\title{PROPERTIES OF SOLAR EPHEMERAL REGIONS AT THE EMERGENCE STAGE}

\author{SHUHONG YANG\altaffilmark{1}, JUN ZHANG\altaffilmark{1}}

\altaffiltext{1}{Key Laboratory of Solar Activity, National
Astronomical Observatories, Chinese Academy of Sciences, Beijing
100012, China; shuhongyang@nao.cas.cn; zjun@nao.cas.cn}

\begin{abstract}

For the first time, we statistically study the properties of
ephemeral regions (ERs) and quantitatively determine their parameters at the emergence
stage based on a sample of 2988 ERs observed by the \emph{Solar
Dynamics Observatory}. During the emergence process, there are
three kinds of kinematic performances, i.e., separation of dipolar
patches, shift of ER's magnetic centroid, and rotation of ER's axis.
The average emergence duration, flux emergence rate, separation
velocity, shift velocity, and angular speed are 49.3 min, 2.6
$\times$ 10$^{15}$ Mx s$^{-1}$, 1.1 km s$^{-1}$, 0.9 km s$^{-1}$,
and 0$\degr$.6 min$^{-1}$, respectively. At the end of emergence,
the mean magnetic flux, separation distance, shift distance, and
rotation angle are 9.3 $\times$ 10$^{18}$ Mx, 4.7 Mm, 1.1 Mm, and
12$\degr$.9, respectively. We also find that the higher the ER
magnetic flux is, (1) the longer the emergence lasts, (2) the higher
the flux emergence rate is, (3) the further the two polarities
separate, (4) the lower the separation velocity is, (5) the larger
the shift distance is, (6) the slower the ER shifts, and (7) the
lower the rotation speed is. However, the rotation angle seems not
to depend on the magnetic flux. Not only at the start time, but also
at the end time, the ERs are randomly oriented in both the northern
and the southern hemispheres. Besides, neither the anticlockwise
rotated ERs, nor the clockwise rotated ones dominate the northern or
the southern hemisphere.

\end{abstract}


\section{INTRODUCTION}

The magnetic flux of dipolar regions emerging from below the solar
surface ranges from less than 10$^{18}$ Mx to more than 10$^{23}$
Mx. The small-scale dipolar regions with short lifetimes were named
ephemeral regions (ERs) and their maximum total flux is $\sim$
10$^{20}$ Mx and the typical lifetime is 1--2 days, as found in the
early study by Harvey \& Martin (1973). Schrijver et al. (1998)
examined the observations from the \emph{Solar and Heliospheric
Observatory} (\emph{SOHO}) and noticed that the mean total unsigned
flux per ER is 1.3 $\times$ 10$^{19}$ Mx. Using the \emph{Hinode}
magnetograms, Wang et al. (2012) quantified the characters of
intranetwork (IN) ERs. Their results reveal that the IN ERs have a
lifetime of 10--15 min and a total maximum unsigned magnetic flux of
the order of 10$^{17}$ Mx. To divide active regions and ERs,
a size limitation of the dipolar area of about 2.5
deg$^{2}$ can be used (Harvey 1993). In a study of Hagenaar et al.
(2003), the ERs are defined as dipoles with total unsigned flux less
than 3 $\times$ 10$^{20}$ Mx.

In the quiet Sun, ERs emerge continuously and thus replenish the
loss of magnetic flux caused by the dispersion and cancellation
(Schrijver et al. 1998). In the initial emergence phase which lasts
about 30 min, the ER's opposite polarity patches rapidly separate up
to about 7 Mm with a velocity of about 4 km s$^{-1}$ (Schrijver et
al. 1998). Then the dipolar patches drift with the supergranular
flow, slowing down to about 0.4 km s$^{-1}$ (Schrijver et al. 1998;
Simon et al. 2001; Priest et al. 2002). With the magnetograms from
the \emph{Solar Dynamics Observatory} ({\it SDO}; Pesnell et al.
2012), Zhao \& Li (2012) studied the properties of 50 ERs.
They selected the ERs, which are
isolated and near the disk center, and have a continuous emergence
phase longer than at least one hour, as the sample. Their results
show that the emerged flux has a range of (0.44--11.2) $\times$
10$^{19}$ Mx, and the emergence duration ranges from 1 hr to 12 hr.
For the IN ERs, Wang et al. (2012) noticed that, during magnetic
flux emergence, most of them display a rotation of the axis and the
rotation angles are more than 10$\degr$.

Although ERs have been extensively studied (Martin \& Harvey 1979;
Martin 1988; Webb et al. 1993; Chae et al.2001; Hagenaar 2001;
Hagenaar et al. 2008), their origin is still under debate. Some
studies (Harvey et al. 1975; Hoyng 1992) suggest that ERs may be the
small-scale tail of a wide spectrum of magnetic activity and also
generated by the global dynamo, which has been commonly considered
as the production mechanism of active regions (Kosovichev 1996;
Dikpati \& Gilman 2001; Mason et al. 2002). This means that ERs are
speculated to come from the bottom of convective zone. However, many
authors argued that, ERs are generated in local turbulent
convection, i.e., by a local dynamo populating everywhere near the
solar surface (Nordlund 1992; Cattaneo 1999; Hagenaar et al. 2003;
Stein et al. 2003). Besides the above models, some people (Nordlund
1992; Ploner et al. 2001) also proposed another possible origin,
i.e., recycle of magnetic flux from decayed and dispersed active
regions.

In a former study (Yang et al. 2012) using observations from the
Helioseismic and Magnetic Imager (HMI; Schou et al. 2012; Scherrer
et al. 2012) onboard {\it SDO}, we have reported that ERs can be
classified into two types: normal ERs and self-canceled ones. Both
of them have the same early evolution process: emerging and growing
with separation of the opposite polarities. After that, the dipolar
patches of normal ERs cancel or merge with the surrounding magnetic
fields, while for self-canceled ones, a part of magnetic fields with
opposite polarities move back, meet together, and cancel with each
other gradually, performing a behavior termed ``self-cancellation."
Considering the same emergence process of the normal ERs and the
self-canceled ones, we can combine them together and investigate
their properties at the emergence stage.

The properties of ERs during emergence process are helpful
for us to provide the necessary parameters for numerical simulation
and to understand the essence of ERs, e.g.,
to know the ratio of emergence duration to lifetime.
In previous studies, only several ER parameters at the emergence stage are
roughly determined just based on small samples (e.g., Schrijver et al. 1998;
Wang et al. 2012; Zhao \& Li 2012).
Hagenaar (2001) studied some basic properties
(including magnetic flux, emergence rate, separation distance, and separation velocity)
of 38000 auto-detected ERs, however the data she adopted are
\emph{SOHO} Michelson Doppler Imager (MDI)
magnetograms with low tempo-spatial resolution, so ERs are still worth studying with
high quality observations (e.g., \emph{SDO}$/$HMI magnetograms).
So far no study has quantitatively determined the parameters of ERs,
such as emergence duration, flux emergence rate, area, flux density, separation
distance, separation velocity, shift distance, shift velocity,
orientation, especially rotation parameters, and
relationships with ER magnetic flux at the definitely defined emergence stage
based on a large sample.

This paper aims to statistically study for the first time the properties of ERs
and quantitatively determine their parameters at the emergence stage
(the start time and the end time are definitely defined by us)
with a large sample of ERs observed by \emph{SDO}.
In Section 2, we describe the
observations and data analysis, and in Section 3, we present the
parameters and behaviors of ERs. The conclusions and discussion are
given in Section 4.

\section{OBSERVATIONS AND DATA ANALYSIS}

The \emph{SDO}/HMI uninterruptedly observes the Sun and records the
line-of-sight magnetic fields with a cadence of 45 s in 6173 {\AA}
line. The full-disk magnetograms have a pixel size of 0$\arcsec$.5
and are free from atmospheric distortions. These advantages are
helpful for us to statistically study the properties of ERs. In this
study, we use a sequence of line-of-sight magnetograms observed by
the HMI in a four-day period (from 2010 June 11 12:00 UT to June 15
12:00 UT). In addition, the 171 {\AA} image observed
by the Atmospheric Imaging Assembly (AIA; Lemen et al. 2012)
at 12:00 UT on June 13 is also adopted.

Since the noise is large for the area far away from the disk center,
only the pixels with heliocentric angle $\alpha$ smaller than
60{\degr} are considered (outlined by the red circle in Figure 1).
All the magnetograms are derotated differentially to a reference
time, June 13 12:00 UT. The blue curve figures out our target where
$\alpha$ $<$ 60{\degr} through the observation period of four days.
In the derotated magnetograms, the area \emph{S} of one pixel is
corrected to $S/\cos(\alpha_{0})$, where $\alpha_{0}$ is the
heliocentric angle of the pixel. While the magnetic flux density $B$
is converted to $B/\cos(\alpha_{1})$, where $\alpha_{1}$ is the
heliocentric angle at the time when it was observed.

We produce movies with the HMI magnetograms firstly, and then try
our best to carefully examine them by eyes. ERs are identified as
dipolar patches with opposite polarities emerge simultaneously or
one following the other, and then grow and separate. Each ER is
identified and tracked by eyes, which makes our results are very
reliable. We define the time when both the positive and the negative patches
are detected as the start time (\emph{t$_{0}$}) of the emergence. If
the strength of a magnetic patch exceeds the noise level, 10.2 Mx
cm$^{-2}$ (Liu et al. 2012), we consider it as being detected. When
the total unsigned magnetic flux of the two polarities reaches the
maximum, the time is defined as the end time (\emph{t$_{1}$}) of the
emergence stage. The ER area is calculated from the pixels occupied
by the magnetic fields stronger than the noise level. The separation
distance is measured along the great circle on the solar surface
between the magnetic centroids of positive and negative patches.
Similar to the separation distance, the shift distance is also
calculated on the sphere, but between the magnetic centroid of the
entire ER at \emph{t$_{0}$} and that at \emph{t$_{1}$}. The
definition of ER's orientation angle $\gamma$ is illustrated in the
inserted circular area in Figure 8(a). ``P" and ``N" represent the
magnetic centroids of ER's positive and negative polarities,
respectively. $\gamma$ is defined as the azimuth of ``P--N" in the
heliographic coordinate, where the west is 0$\degr$ and the north is
90$\degr$, instead of in the Cartesian coordinate. The range of
$\gamma$ is (0$\degr$, 360$\degr$). The flux emergence rate,
separation velocity, shift velocity, and angular speed of rotation
are the average values during emergence.


\section{RESULTS}

\subsection{Latitudinal distribution and magnetic flux of ERs}

In the target delineated by the blue curve (see Figure 1), we
identified 2988 ERs in all. After projection correction, the real
area of our target outlined by the blue curve in Figure 1 is 8.0
$\times$ 10$^{5}$ Mm$^{2}$, and thus the average number density in
the whole target is 9.32 $\times$ 10$^{-4}$ day$^{-1}$ Mm$^{-2}$.
This value is larger than that
(8.7 $\times$ 10$^{-4}$ day$^{-1}$ Mm$^{-2}$) determined by Wang (1988)
and that (6.7 $\times$ 10$^{-4}$ day$^{-1}$ Mm$^{-2}$) determined by Martin (1989),
while it is smaller than those
(17 $\times$ 10$^{-4}$ day$^{-1}$ Mm$^{-2}$,
27 $\times$ 10$^{-4}$ day$^{-1}$ Mm$^{-2}$,
71 $\times$ 10$^{-4}$ day$^{-1}$ Mm$^{-2}$)
determined by Chae et al. (2001), Harvey (1993), and Hagennar (2001), respectively.
The different values of average number density are marked with dashed horizontal lines in Figure 2(a).
In our opinion, different data from different instruments and different sample sizes may lead to the difference of the results.
The latitudinal distribution of these ERs is presented in Figure
2(a). The ERs are not uniformly distributed in the range of
(S60$\degr$, N60$\degr$). There are two regions with larger number
density: one is located around S15$\degr$ and the other one around
N25$\degr$.
This appearance was also presented in the study of Hagennar et al. (2003).
The number densities at these two regions exceed 10
$\times$ 10$^{-4}$ day$^{-1}$ Mm$^{-2}$, higher than the density
(8.8 $\times$ 10$^{-4}$ day$^{-1}$ Mm$^{-2}$) at the equatorial
region between S10$\degr$ and N10$\degr$. The regions with latitude
higher than 40$\degr$ have much lower density, only about (5 $\sim$
6) $\times$ 10$^{-4}$ day$^{-1}$ Mm$^{-2}$. The probability density
function (PDF) of the ERs with a bin size of 0.6 $\times$ 10$^{18}$
Mx is plotted in Figure 2(b). We can see that there is a PDF peak at
3.6 $\times$ 10$^{18}$ Mx, and the mean magnetic flux is 9.27
$\times$ 10$^{18}$ Mx.
This value is close to the results of
Hagenaar (2001; 1.13 $\times$ 10$^{19}$ Mx),
Schrijver et al. (1998; 1.3 $\times$ 10$^{19}$ Mx), and
Wang (1988; 1.5 $\times$ 10$^{19}$ Mx), but it is not well consistent with
several other results (2.5 $\times$ 10$^{19}$ Mx, Martin 1989; 2.8 $\times$ 10$^{19}$ Mx, Chae et al. 2001;
3.3 $\times$ 10$^{19}$ Mx, Harvey 1993; 3.9 $\times$ 10$^{19}$ Mx, Zhao \& Li 2012),
especially the very early study ($\sim$ 10$^{20}$ Mx) by Harvey \& Martin (1973).
The mean values reported by different studies are marked with dashed vertical lines in Figure 2(b).
This difference may be due to the better sensitivity
and higher spatial resolution of \emph{SDO}/HMI.
Although Zhao \& Li (2012) also used HMI data, they only selected large ERs,
so they obtained a higher ER magnetic flux.
Wang et al. (2012) investigated ERs using \emph{Hinode} data with high spatial resolution,
however they just focused on IN ERs. They found that the IN ERs have magnetic fluxes from
several 10$^{16}$ Mx to 1.5 $\times$ 10$^{18}$ Mx, and the average flux is about 0.8 $\times$ 10$^{18}$ Mx,
more than one order of magnitude smaller than our result.
The magnetic flux of the ERs spreads from
smaller than 10$^{18}$ Mx to larger than 10$^{20}$ Mx, but none of
them is larger than 3 $\times$ 10$^{20}$ Mx, the upper limitation of
ER flux defined by Hagenaar et al. (2003).

\subsection{Separation and growth of ER's opposite polarities}

The most conspicuous performance of ERs at the emergence stage is
the separation and growth of dipolar patches. Figure 3 displays
three magnetograms showing the separation and growth of the opposite
polarities of an ER on 2010 June 14. The start time \emph{t$_{0}$}
of the ER emergence was 04:23 UT. At that time, both of the positive
(white patch with blue contour) and negative (black patch with red
contour) polarities were strong enough to be detected (panel (a)).
The total unsigned magnetic flux of the ER was 4.3 $\times$
10$^{17}$ Mx, the area occupied by the two patches was 2.7 Mm$^{2}$,
and the mean absolute flux density was 15.9 Mx cm$^{-2}$.  The
distance (marked by the green curve) between the two magnetic
centroids (indicated by the blue and red plus signs) was 1.9 Mm.
Then 6 min later, the ER area expanded conspicuously, and the
strength was much stronger compared with the initial appearance at
the start time (panel (b)).
At 04:32 UT, the total unsigned magnetic flux of the ER reached the
maximum, 2.7 $\times$ 10$^{18}$ Mx. According to the definition, it
was the end time \emph{t$_{1}$}, indicating the end of emergence.
Thus, the duration of emergence was 9 min. The area expanded to 12.3
Mm$^{2}$, the mean flux density changed to 21.8 Mx cm$^{-2}$ and the
separation distance reached 3.2 Mm. During the emergence period, the
average separation velocity was 2.5 km s$^{-1}$ and the average flux
emergence rate was 4.2 $\times$ 10$^{15}$ Mx s$^{-1}$.

The PDFs of separation distance at \emph{t$_{1}$}, emergence
duration, separation velocity, flux emergence rate, area at
\emph{t$_{1}$}, and flux density at \emph{t$_{1}$} are displayed in
Figure 4. Each of them has a peak value. The separation distance
ranges from 1.3 Mm to 19.2 Mm with a peak at 3.2 Mm, and the mean
value is 4.7 Mm (panel (a)). The peak of the emergence duration is
at 12 min, and the mean duration is 49.3 min (panel (b)). During the
emergence, the separation velocities of ERs vary between 0.03 km
s$^{-1}$ and 5.5 km s$^{-1}$, and their PDF peaks at 0.4 km s$^{-1}$
(panel (c)). The mean separation velocity is 1.1 km s$^{-1}$. The
peak value and mean value of flux emergence rate are at 1.0 $\times$
10$^{15}$ Mx s$^{-1}$ and 2.6 $\times$ 10$^{15}$ Mx s$^{-1}$,
respectively (panel (d)). The areas at the end of emergence have a
distribution peak at 10.0 Mm$^{2}$ and the mean area is 23.1
Mm$^{2}$ (panel (e)). The magnetic flux densities of the ERs are
also determined at the end of emergence. Most ERs are not strong,
only several ten Mx cm$^{-2}$, and the peak value of PDF is at 28.0
Mx cm$^{-2}$, as shown in panel (f). The mean flux density of ERs is
35.3 Mx cm$^{-2}$.
Schrijver et al. (1998) found that the
initial emergence phase lasts about 30 min, the separation distance
is up to about 7 Mm, and the separation velocity is about 4 km
s$^{-1}$ (marked with dashed vertical lines labeled with ``2" in Figure 4).
In the study of Hagenaar (2001), the average distance between the opposite
polarities is 8.9 Mm and the expanding velocity is 2.3 km s$^{-1}$ (marked with lines ``4").
As reported by Chae et al. (2001) and Harvey \& Harvey (1976),
the values of average separation of fully developed ERs are 7.4 Mm
(marked with line ``3") and 2$-$3 Mm (line ``6"), respectively.
The separations of IN ERs obtained by Wang et al. (2012) are 3$\arcsec$$-$4$\arcsec$ and
the average value is 3$\arcsec$.3, i.e., 2.4 Mm (line ``7").
At the very beginning phase of emergence, the separation velocity determined by Title (2000)
is of the order of 5 km s$^{-1}$ (line ``8").
The mean flux emergence rates determined by Hagenaar (2001), Zhao \& Li (2012),
Harvey \& Harvey (1976), and Wang (1988; line ``9") are
1.6 $\times$ 10$^{15}$ Mx s$^{-1}$,
2.31 $\times$ 10$^{15}$ Mx s$^{-1}$,
3.4 $\times$ 10$^{15}$ Mx s$^{-1}$,
and 2.2 $\times$ 10$^{15}$ Mx s$^{-1}$, respectively.
Their results are generally consistent with ours. However, there
are also some differences which may be due to different observations.
According to the results (vertical lines labeled with ``5")
of Zhao \& Li (2012), the values of some parameters (such as unsigned flux,
duration, distance, and area) are consistent with these with large
magnetic flux in our study, since they selected their sample with
three criterions, and thus small ERs were ignored.
According to the results of Harvey (1993), the lifetime of ERs is 4.4 hr.
The emergence duration determined in the present study is about 50 min,
so the birth stage is only about 0.19 of the lifetime.

Figure 5 shows the relationships between the parameters mentioned
above and the total unsigned magnetic flux of ERs. The red dots are
the scatter plots of all the ERs. In order to well display the
relationships, the data are processed with a ``sort-group" method
(Zhao et al. 2009). Firstly, the data are sorted in ascending order
according to the total unsigned flux. Then each 300 ERs are grouped
into one group, and 10 groups are obtained in all. Finally, the
parameters are correlated with the magnetic flux, and the values of
ten groups are plotted (marked by the blue symbols in each panel).
Each error bar represents the standard deviation of the
corresponding group data. We can see that the separation distance
(panel (a)), emergence duration (panel (b)), flux emergence rate
(panel (d)), area (panel (e)), and flux density (panel (f)) are
positively correlated with the magnetic flux. While the separation
velocity has a negative correlation with the magnetic flux (panel
(c)). The variation trends indicate that, the higher the total
unsigned magnetic flux is, (1) the further the two polarities
separate, (2) the longer the emergence last, (3) the slower the
separation is, (4) the higher the flux emergence rate is, (5) the
larger the area is, and (6) the stronger the ER field is.
Harvey \& Harvey (1976) have found that the flux emergence
rate is large for the larger ERs, and Zhao \& Li (2012) showed that the emergence
duration and flux growth rate are positively correlated to the total
emerging flux, which are supported by our results (see Figures 5(b)
and 5(d)).

\subsection{Shift of ER's magnetic centroid}

We also compute the magnetic centroid of each ER using its total
unsigned magnetic flux. During the emergence, the magnetic centroid
does not keep stable, and there exists a shift on the solar surface.
Figure 6 shows the evolution of an ER as an example to illustrate
this kind of movement. At 00:44 UT on June 12, the ER emergence
began (panel (a)). The ER centroid was marked by the green plus
symbol, and it was located at ($-$6.$\arcsec$5, 19.$\arcsec$1). At
00:53 UT, the location of magnetic centroid moved to
($-$7.$\arcsec$3, 18.$\arcsec$6), indicating that the ER moved in a
south-east direction (panel (b)). When the emergence ended, the ER
reached to ($-$8.$\arcsec$2, 17.$\arcsec$4) (panel (c)). The shift
movement relative to the heliospheric grids can also be easily
found. From \emph{t$_{0}$} to \emph{t$_{1}$}, the centroid shifted
1.76 Mm in 54 min with an average shift velocity of 0.54 km
s$^{-1}$.

For the shift distance and shift velocity of ERs, their PDFs and
their relationships with the total unsigned magnetic flux are
presented in Figure 7. The shift distance is no more than several
Mm, and the mean distance is 1.1 Mm while the peak value is at 0.4
Mm (panel (a)). The PDF of shift velocity peaks at 0.2 km s$^{-1}$
and the mean velocity is 0.9 km s$^{-1}$ (panel (b)). Similar to the
relationship between the separation distance and the total unsigned
magnetic flux shown in Figure 5(a), there is also a positive
correlation for the shift distance as seen in Figure 7(c). The
increase trend indicates that the higher the total magnetic flux is,
the larger the shift distance is. The shift velocity is negatively
correlated with the magnetic flux (panel (d)), indicating that the
higher the total magnetic flux is, the slower the ER's centroid
shifts.

\subsection{Orientation and rotation of ERs}

Besides the separation movement and shift movement, ERs also rotate
during the emergence, leading to the change of orientation. An
example of ER rotation is given in Figure 8.
At 10:02 UT on June 13, both the positive and negative polarities of
the ER appeared. The initial orientation is shown in panel (a). The
negative patch was located at the north-east of the positive patch,
and the orientation was 113.$\degr$1. Then the axis of the ER
rotated clockwise, and 6 min later, the orientation changed to
101.$\degr$1 (panel (b)), much smaller than the initial angle. The
emergence went on, and the clockwise rotation also did not stop. At
the end of of emergence, there was a significant change of the ER
appearance (panel (c)). The orientation became 93.$\degr$1, which
means the ER rotated clockwise 20.$\degr$0 with an absolute average
angular speed of 2.$\degr$2 min$^{-1}$ at the emergence stage.

Due to the rotation, the orientations of ERs at \emph{t$_{0}$} and
at \emph{t$_{1}$} are different. Figure 9 displays the
spatial distribution of the orientations at the start time (panel
(a)) and the end time (panel (b)) of flux emergence. In the whole
target, the orientations both at \emph{t$_{0}$} and \emph{t$_{1}$}
are randomly distributed, but in some sizable (100$\arcsec$ $\times$
100$\arcsec$) areas, ERs are generally ordered. For example, as
shown in panel (c1), the orientations of ERs outlined by the ellipse
are mainly from the positive to the negative polarities (see panel
(c2)). The inserted colorful image in panel (c2) is the AIA 171
{\AA} observation and displays the overlying coronal loops
(emphasized with dotted green curves) between positive and negative
magnetic fields. In panel (d1), the general pointing directions of
the ERs located within the green circle are from northwest to
southeast, consistent with the alignment of positive and negative
polarities of the large-scale background fields (see panel (d2)).
The black circle in panel (d1) contains ERs which mainly align in
the southeast--northwest direction, also the direction of background
fields, as shown in panel (d2).
These results imply that, in some sizable areas,
ER orientation may depend on the large-scale magnetic configuration.

The histograms of the orientations at \emph{t$_{0}$} and those at
\emph{t$_{1}$} are presented in the left and the right columns in
Figure 10, respectively. The histograms are shown in angular
representation with a bin size of 45$\degr$, and the ERs located in
the northern and southern hemispheres are separately considered. At
the start time of emergence, the initial orientations are
essentially randomly distributed for both the northern (panel (a))
and southern (panel (b)) ERs. At the end of emergence stage, the
orientations after rotation are still basically randomly oriented in
the northern hemisphere, also in the southern hemisphere.

Figure 11 shows the PDFs of rotation angle and angular speed, and
their relationships with the magnetic flux. If one ER rotates
anticlockwise, its rotation angle and angular speed are assigned a
plus, and if it rotates clockwise, a minus is assigned. The PDF of
rotation angle have a general balance between the positive and
negative values, and the peak is at zero (panel (a)). The mean value
of absolute rotation angles is 12$\degr$.9 (marked with line ``1").
This value agrees with that ($>$ 10$\degr$; marked with line ``2") of IN ERs reported
by Wang et al. (2012).
The data of angular speed also peak at zero, and the mean absolute angular speed is
0$\degr$.6 min$^{-1}$ (panel (b)). The plots in panel (c) show that
there is no close relationship between absolute rotation angle and
magnetic flux, i.e., the rotation angle does not depend on the
magnetic flux. The variation of absolute angular speed with magnetic
flux is presented in panel (d). The decrease trend reveals that the
higher the magnetic flux is, the lower the absolute angular speed
is.

Some ERs rotate clockwise, while some anticlockwise. In Figure 1,
the red dots represent the ERs with anti-clockwise rotation, while
the blue ones represent the ERs with clockwise rotation. In order to
examine if there exists a hemisphere rule for the ER rotation, the
ERs at different latitudes should be considered separately. We
define an imbalance parameter $\rho$ to describe the number
imbalance between the anticlockwise and the clockwise rotated ERs.
$\rho$ is defined as
\begin{equation}
\rho =\frac{ N_{anticlockwise} - N_{clockwise} }{ N_{anticlockwise} + N_{clockwise} },
\end{equation}
where ``N" is the ER number. The parameter $\rho$ as a function of
latitude is shown in Figure 12. It shows that there is no
significant domination of anticlockwise rotation or of clockwise
rotation at different latitudes. The average value of $|\rho|$ is
about 0.03 and the maximum value is smaller than 0.07.

\section{CONCLUSIONS AND DISCUSSION}

In this study, we have statistically investigated the properties of ERs.
Based on almost three thousands of ERs, we quantitatively determine
their parameters at the emergence stage for the first time.
During the emergence process, there are three kinds of kinematic
performances, e.g., separation of dipolar patches, shift of ER's
magnetic centroid, and rotation of ER's axis. Several parameters,
e.g., duration of emergence, flux emergence rate, and separation
velocity, are measured. At the end of emergence, six parameters,
i.e., magnetic flux, area, flux density, separation distance, shift
distance, and rotation angle are also determined.
Moreover, we find that the higher the ER magnetic flux is, (1) the
further the two polarities separate, (2) the longer the emergence
lasts, (3) the slower the separation is, (4) the higher the flux
emergence rate is, (5) the larger the area is, (6) the stronger the
ER field is, (7) the larger the shift distance is, (8) the slower
the ER's centroid shifts, and (9) the lower the absolute angular
speed is.
In addition, we notice that the regions with locations around
S15$\degr$ and around N25$\degr$ have larger number density.

There are at least three kinds of convective cells according to
their sizes, i.e., granulation, mesogranulation, and
supergranulation (Simon \& Leighton 1964; Rast 2003). The horizontal
velocity of supergranular flows is determined to be about 0.3$-$0.5
km s$^{-1}$ (Leighton et al. 1962; Simon \& Leighton 1964). Magnetic
elements which emerge within the supergranular cells are advected
toward to the supergranular borders with a velocity of about 0.4 km
s$^{-1}$ (Zhang et al. 1998). However, as revealed by our results
and also observed by Schrijver et al. (1998), the separation
velocity of dipolar patches at the emergence stage is much larger.
We suggest that, when magnetic flux tubes rise from below the
photosphere, the $\Omega$$-$shaped tubes are mainly affected by the
buoyant force and can be ejected into the atmosphere of the Sun with
a high velocity, leading to a rapid separation of the positive and
negative polarities. After the emergence stage, magnetic patches are
mainly driven by horizontal supergranular flows, and thus the
separation slows down. We also find that the average shift velocity
(0.9 km s$^{-1}$) of the ERs is larger than that of the
supergranular flows. It may be due to the existence of configuration
asymmetry of ERs. When ERs emerge, the separation of dipolar patches
is not symmetric, so the shifts are observed. Another possible
interpretation is that the planes the flux tubes are located in are
not vertical, i.e., there exits tilt angles relative to the solar
radial direction. As soon as they rise and get through the
photosphere, the flux tubes will become vertical, resulting in the
magnetic centroid sideward shifts (see Figure 6). As shown in this
study, ERs display a rotation of their axes. One possible reason is
that the chaotic convective motions shear and distort the dipolar
patches. Another illustration is that the rotation is caused by the
relaxation of twisted ERs (Patsourakos et al. 2008), implying the
structures of ERs are quite complicate.

The most popular model about the formation of active regions is that
they are generated by a dynamo action at the bottom of the
convection zone, where the tachocline is located (Dikpati \& Gilman
2001; Mason et al. 2002). The tachocline is a transition layer of
solar rotation, from the solid-body rotation of the radiative
interior to the differential rotation of the convection zone
(Kosovichev 1996). The large shear in the tachocline can form and
store large scale fields, which will emerge through the solar
surface due to buoyant force (Parker 1993). Active regions are
aligned generally in the east-west direction with a few degrees
because of the effect of Coriolis force during the emergence of flux
tubes (Hale et al. 1919; Schmidt 1968). But for the ERs in our
study, not only at the start time, but also at the end time, their
orientations are basically randomly oriented in both the northern
and the southern hemispheres (Figure 10). Besides, neither the
anticlockwise rotated ERs, nor the clockwise rotated ones dominate
the northern or the southern hemisphere (Figure 12). The locations
of the ERs spread all over the target from S60$\degr$ to N60$\degr$
(see Figure 1). These results imply that, differing from active
regions, it seems that ER flux is not generated by the
global dynamo, and it may be generated by a local dynamo.
However, small flux systems
are greatly affected by convective motion below the solar
photosphere and will eventually have random orientation when they emerge
even if they may have been generated by the global dynamo. So we cannot exclude
the possibility that they may be generated by the global dynamo. Figure
2(a) shows that, instead of the equatorial or the high latitudinal
regions, the regions at around S15$\degr$ and N25$\degr$ (the
general latitudes of active regions) have larger ER number density,
indicating that the recycle of magnetic flux from decayed and
dispersed active regions may be another origin of the
magnetic flux of ERs.

\acknowledgments { We thank the referee for constructive
comments and Prof. Jingxiu Wang for his helpful suggestions. This
work is supported by the Outstanding Young Scientist Project
11025315, the National Basic Research Program of China under grant
2011CB811403, the National Natural Science Foundations of China
(11203037, 11221063, 11373004, and 11303049), and the CAS Project
KJCX2-EW-T07. The data are used by courtesy of NASA/\emph{SDO} and
the HMI science team.}

{}

\clearpage

\begin{figure}
\centering
\includegraphics
[bb=74 215 488 618,clip,angle=0,scale=1.1] {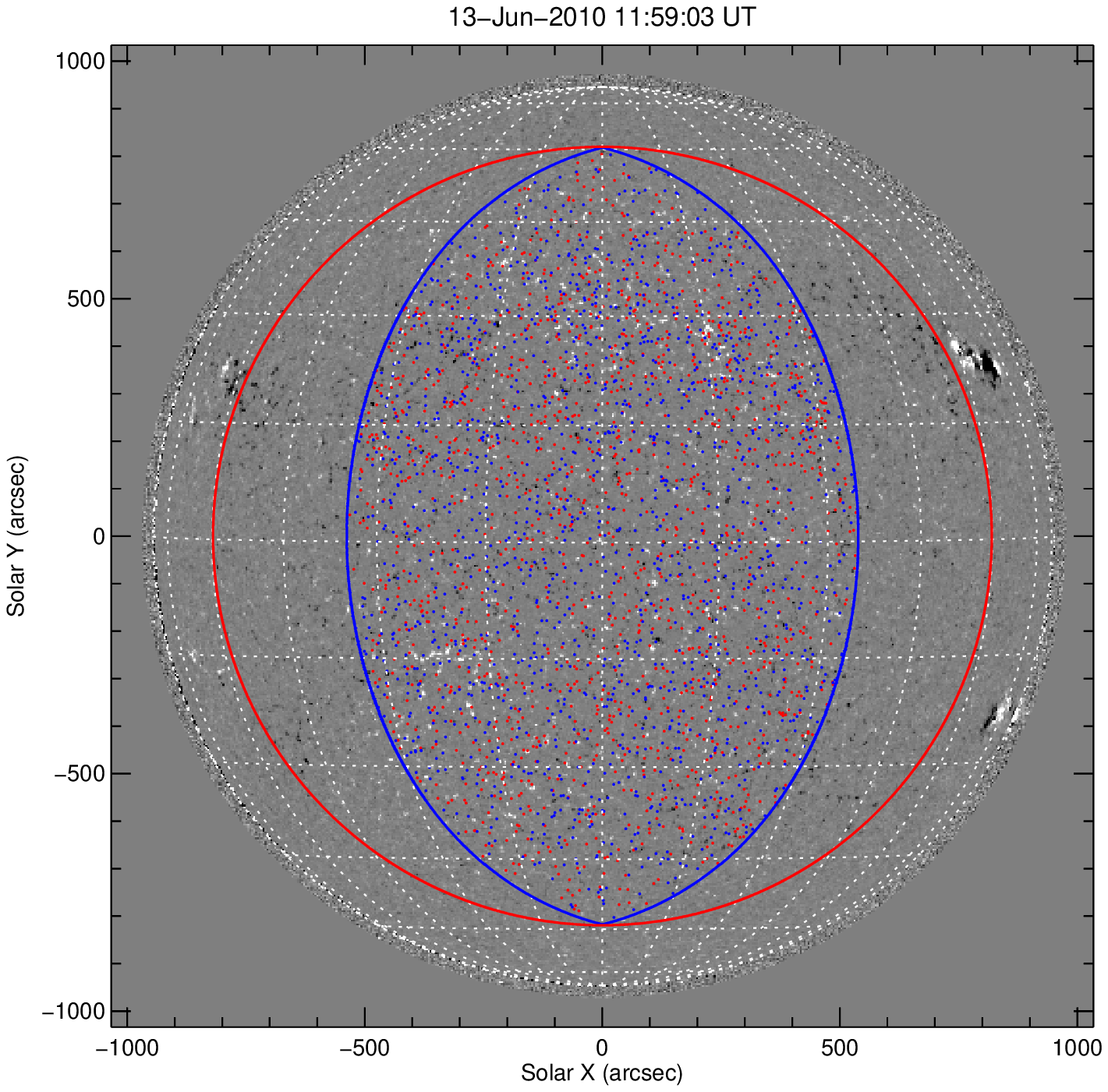}
\caption{Distribution of the ERs. The background is the
line-of-sight magnetogram observed by \emph{SDO}/HMI at 11:59 UT on
2010 June 13. Each red/blue dot represents an ER with
anti-clockwise/clockwise rotation. The red circle is located at the
place with heliocentric angle $\alpha$ of 60{\degr}. The blue curve
figures out the area where $\alpha$ $<$ 60{\degr} in the whole four
days (June 11 12:00 UT -- June 15 12:00 UT). \label{fig1}}
\end{figure}
\clearpage

\begin{figure}
\centering
\includegraphics
[bb=102 123 470 695,clip,angle=0,scale=0.95] {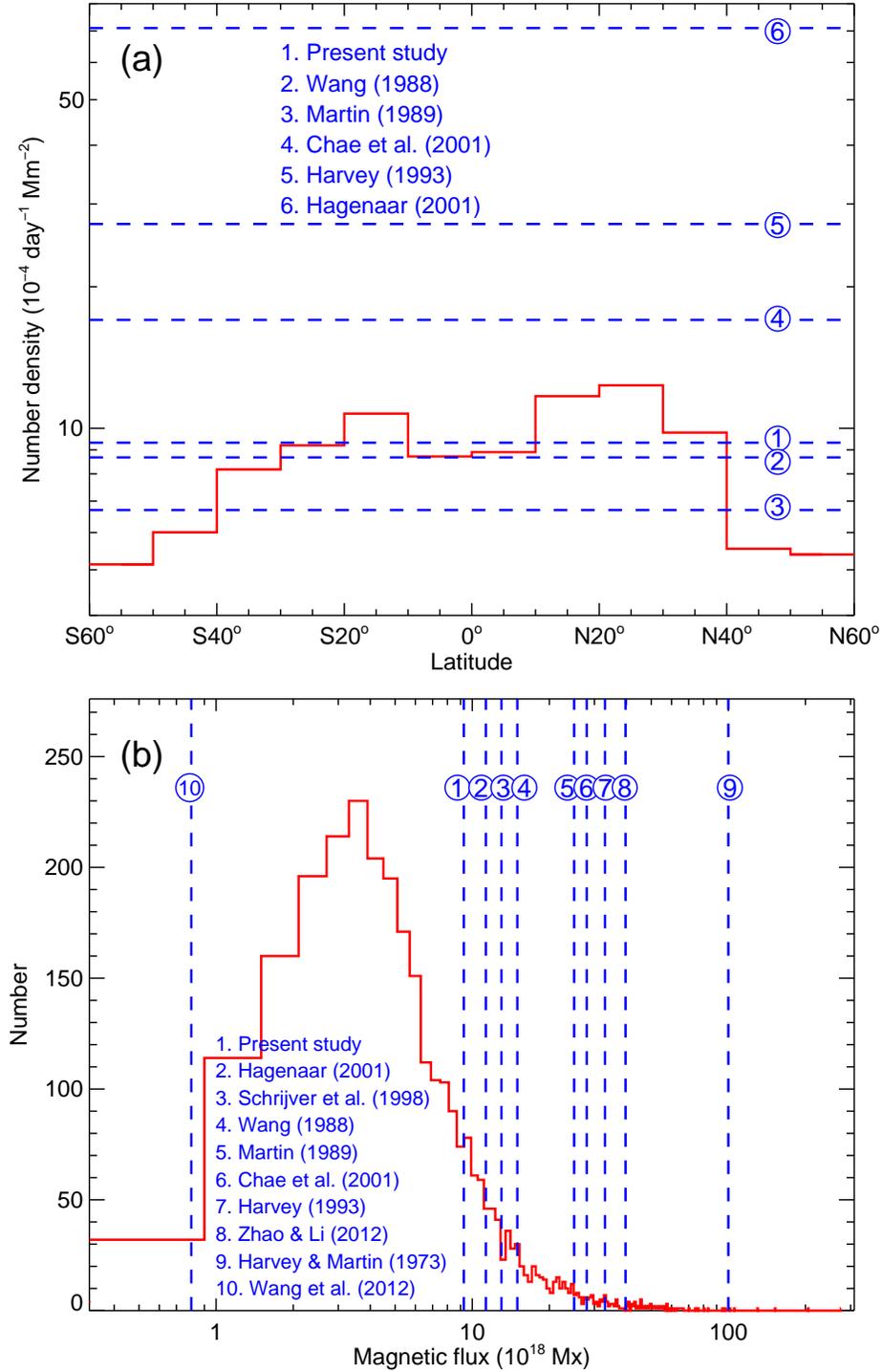}
\caption{\emph{Upper panel}: Number density of ERs as a function of
latitude. \emph{Lower panel}: PDF of total unsigned magnetic flux of
ERs with a bin size of 0.6 $\times$ 10$^{18}$ Mx.
The dashed lines mark different values obtained by different studies. \label{fig2}}
\end{figure}
\clearpage

\begin{figure}
\centering
\includegraphics
[bb=52 332 546 497,clip,angle=0,scale=0.95] {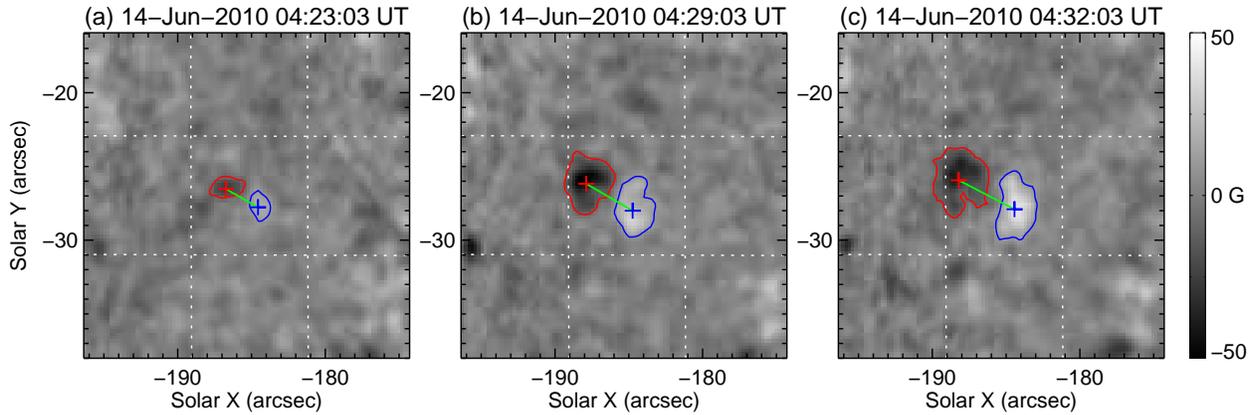}
\caption{Sequence of HMI magnetograms illustrating the separation
and growth of the dipolar patches of an ER at the emergence stage.
The blue and red curves are contours of the ER's positive and
negative polarities at + 10.2 G and -- 10.2 G levels, respectively.
The blue and red plus symbols mark the magnetic centroids of the
positive and negative polarities and the green curves are the
distance between them along the great circle of solar surface. The
dotted curves are the heliographic grids with a grid spacing of
0{\degr}.5. \label{fig3}}
\end{figure}
\clearpage

\begin{figure}
\centering
\includegraphics
[bb=47 253 515 564,clip,angle=0,scale=1] {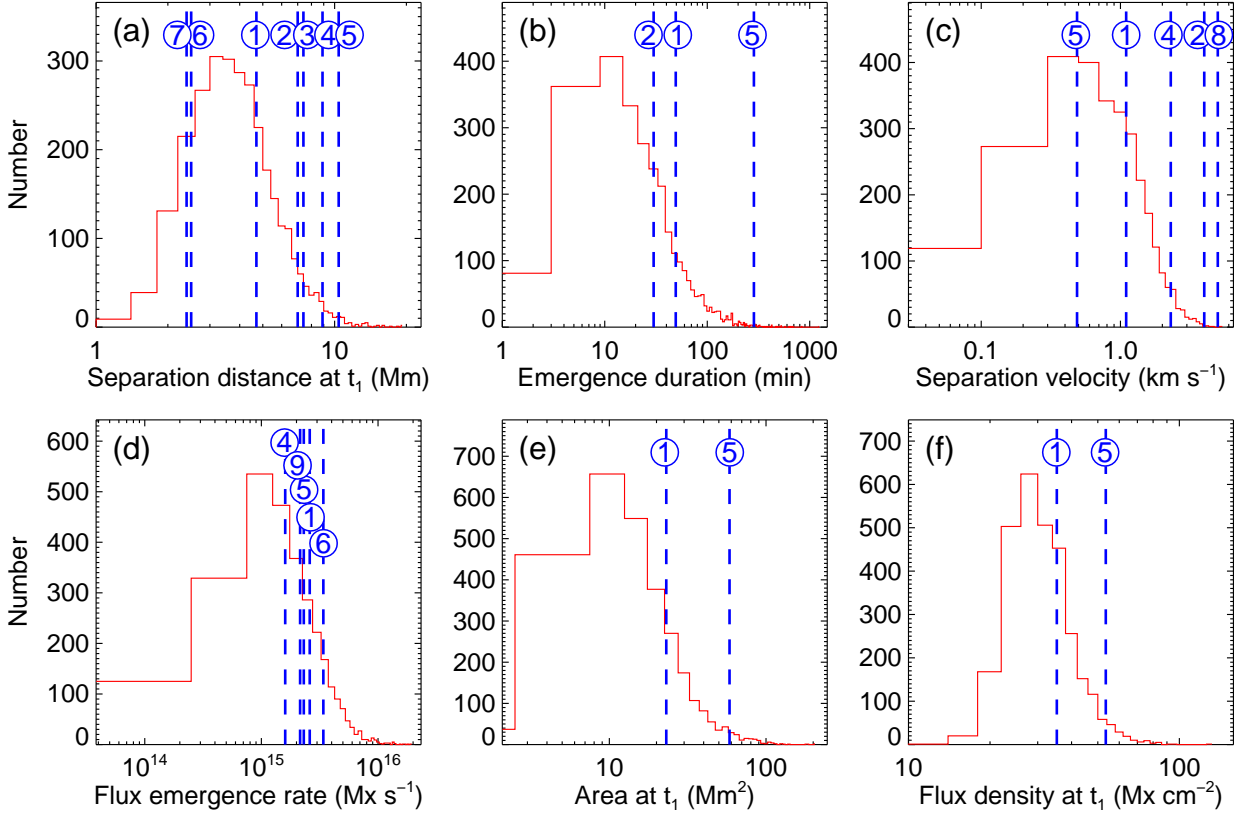} \caption{PDFs of
separation distance (panel (a)) at t$_{1}$, emergence duration
(panel (b)), separation velocity (panel (c)), flux emergence rate
(panel (d)), area (panel (e)) at t$_{1}$, and flux density (panel
(f)) at t$_{1}$.
The values obtained by different studies are marked with dashed vertical lines:
``1" (Present study), ``2" (Schrijver et al. 1998), ``3" (Chae et al. 2001), ``4" (Hagenaar 2001),
``5" (Zhao \& Li 2012), ``6" (Harvey \& Harvey 1976), ``7" (Wang et al. 2012), ``8" (Title 2000), and ``9" (Wang 1988).
\label{fig4}}
\end{figure}
\clearpage

\begin{figure}
\centering
\includegraphics
[bb=47 253 515 564,clip,angle=0,scale=1] {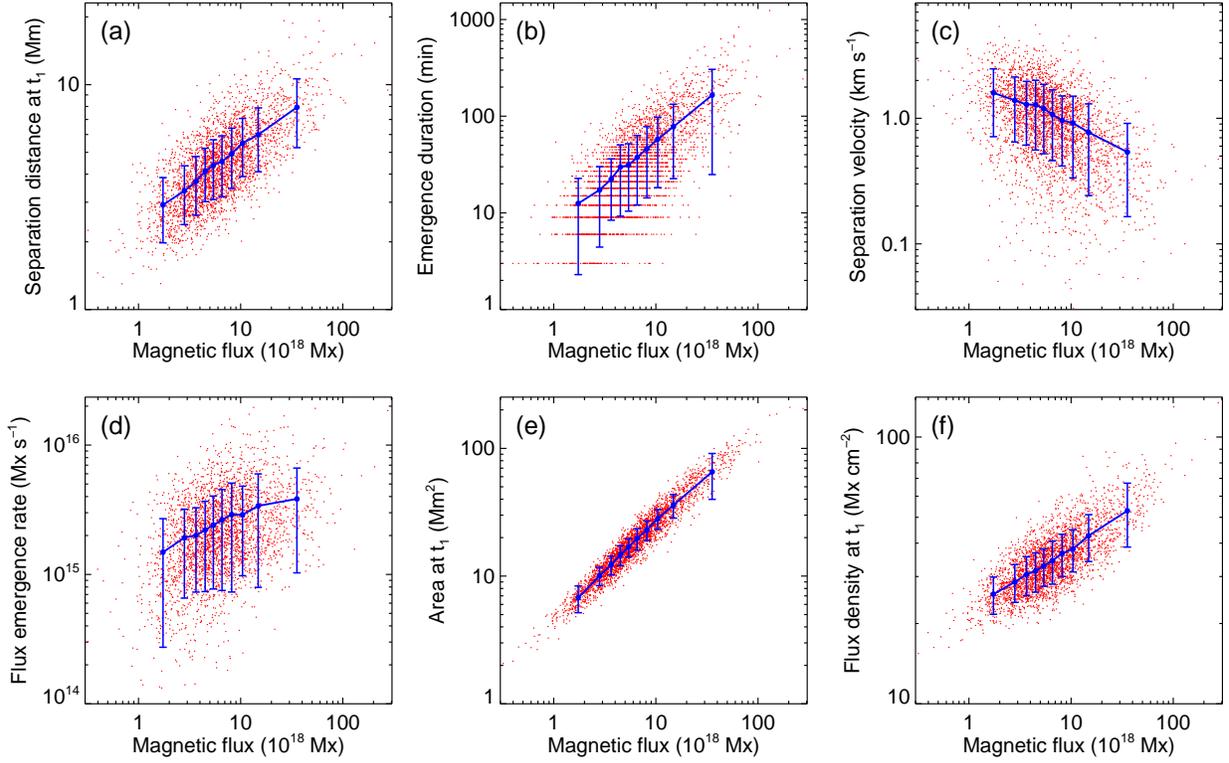} \caption{Scatter
plots of separation distance, emergence duration, separation
velocity, flux emergence rate, area, and flux density versus
magnetic flux of ERs (red symbols) and of sorted and grouped points
(blue symbols) (from panel (a) to panel (f), respectively). Each
error bar represents the standard deviation of the corresponding
group data. \label{fig5}}
\end{figure}
\clearpage

\begin{figure}
\centering
\includegraphics
[bb=52 332 546 497,clip,angle=0,scale=0.95] {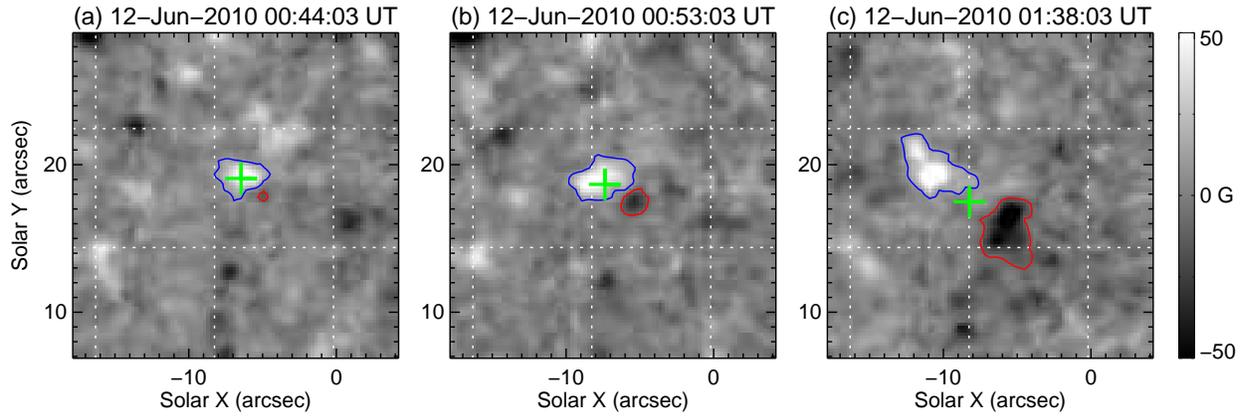}
\caption{Sequence of magnetograms illustrating the shift of magnetic
centroid of an ER at the emergence stage. The blue/red curves are
contours of the ER's positive/negative polarity at +/-- 10.2 G, and
the green plus symbols mark the magnetic centroid of the ER. The
dotted curves are the heliographic grids with a grid spacing of
0{\degr}.5. \label{fig6}}
\end{figure}
\clearpage

\begin{figure}
\centering
\includegraphics
[bb=70 217 484 605,clip,angle=0,scale=1.] {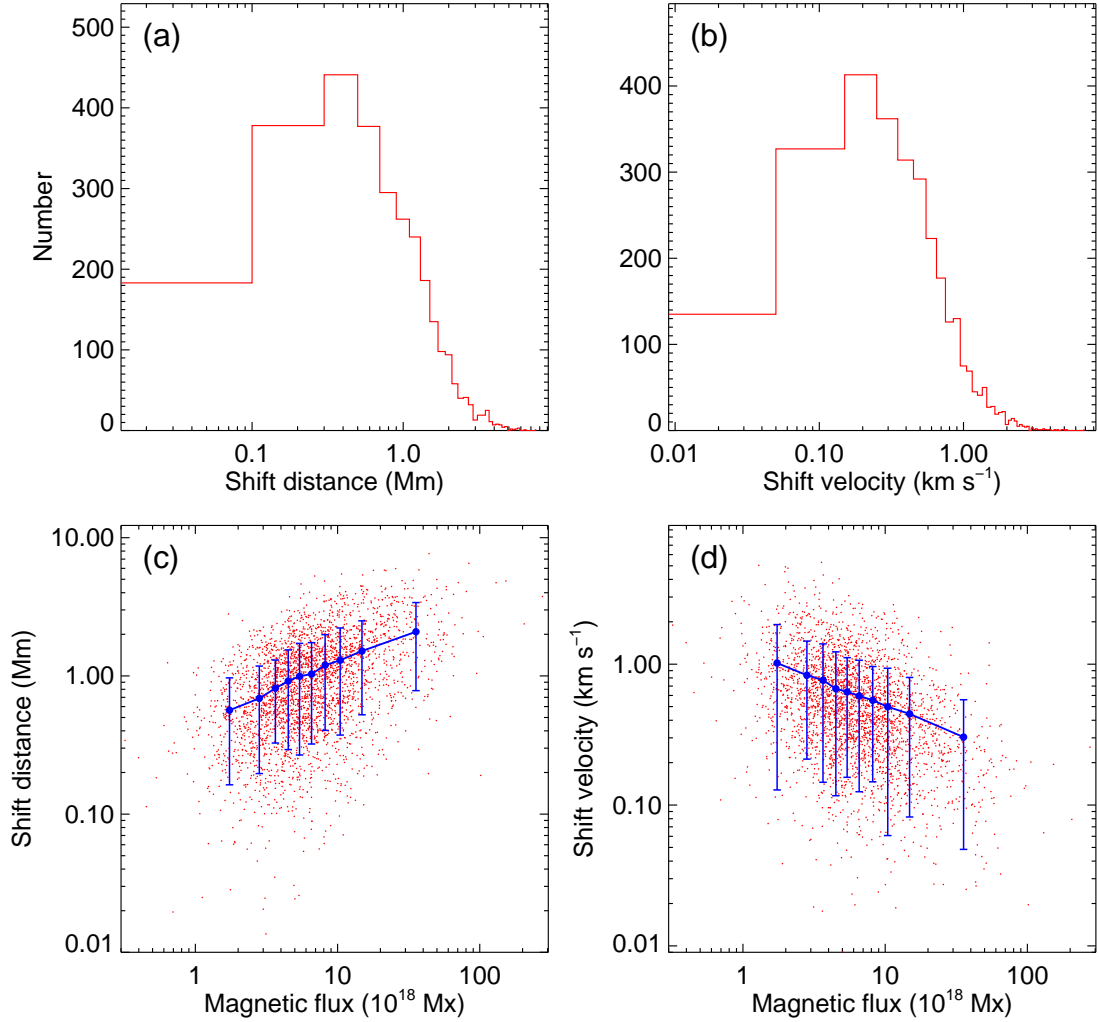}
\caption{\emph{Upper panels}: PDFs of shift distance (panel (a)) and
shift velocity (panel (b)). \emph{Lower panels}: Scatter plots of
shift distance (panel (c)) and shift velocity (panel (d)) versus
magnetic flux of ERs. \label{fig7}}
\end{figure}
\clearpage

\begin{figure}
\centering
\includegraphics
[bb=52 332 546 497,clip,angle=0,scale=0.95] {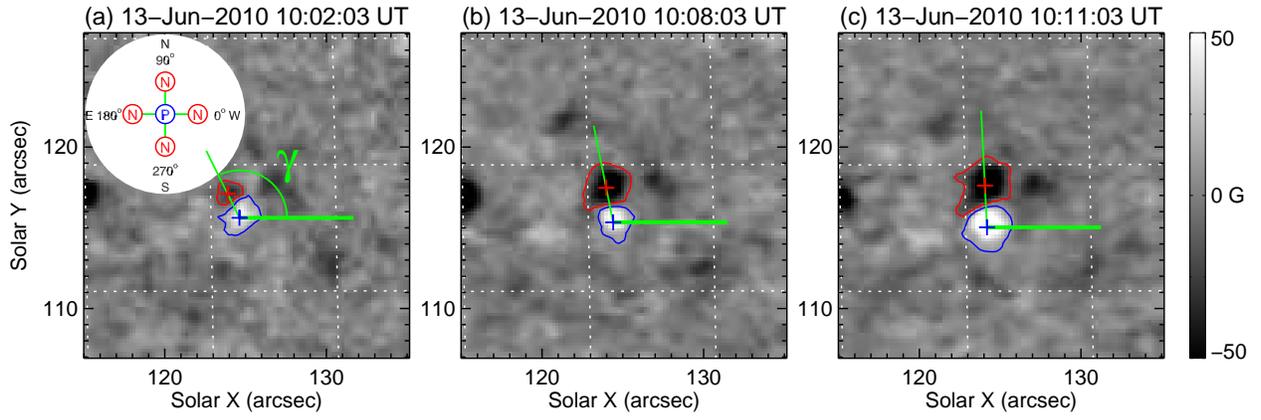}
\caption{Sequence of magnetograms illustrating the rotation of an ER
at the emergence stage. The contours, plus signs, and dotted curves
have the same meanings as those in Figure 3. $\gamma$ is the
orientation which varies between (0{\degr}, 360{\degr}). The
definition of $\gamma$ is displayed in the inserted circular area in
panel (a). ``P" and ``N" represent the positive and negative
polarities, respectively. \label{fig8}}
\end{figure}
\clearpage

\begin{figure}
\centering
\includegraphics
[bb=62 187 503 645,clip,angle=0,scale=1.0] {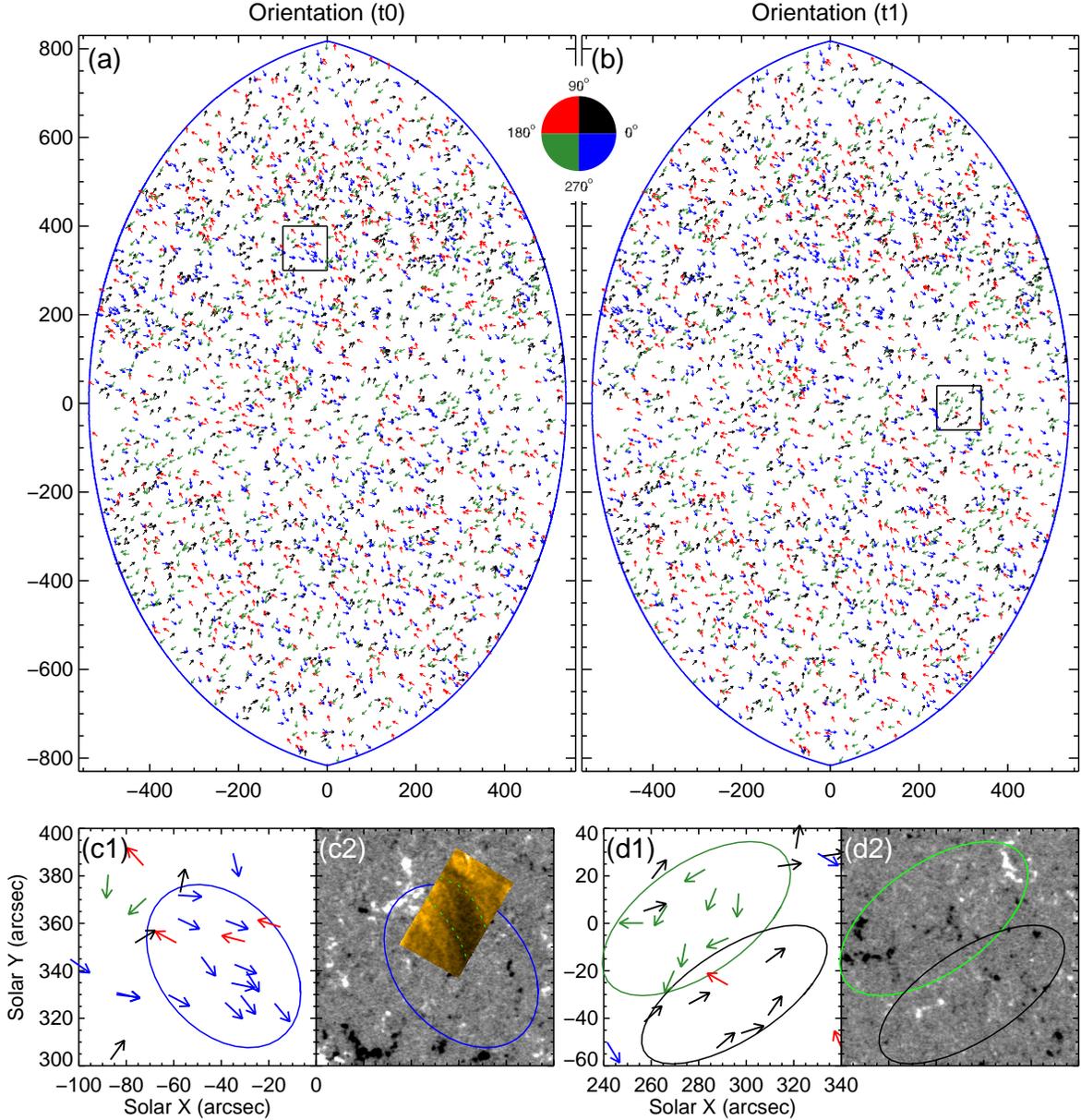}
\caption{Spatial distribution of the ER orientations at the
start time (panel (a)) and the end time (panel (b)) of flux
emergence. Panel (c1) is the enlarged frame outlined by the square
in panel (a), and panel (c2) is the corresponding magnetogram.
Panels (d1) and (d2) are similar to panels (c1) and (c2) but for the
area outlined by the square in panel (b). The inserted colorful
image in panel (c2) is the AIA 171 {\AA} observation and displays
the overlying coronal loops (emphasized with dotted green curves)
between positive and negative magnetic fields. The ellipses outline
the ERs which are generally ordered. \label{fig9}}
\end{figure}
\clearpage

\begin{figure}
\centering
\includegraphics
[bb=135 270 445 571,clip,angle=0,scale=1.2] {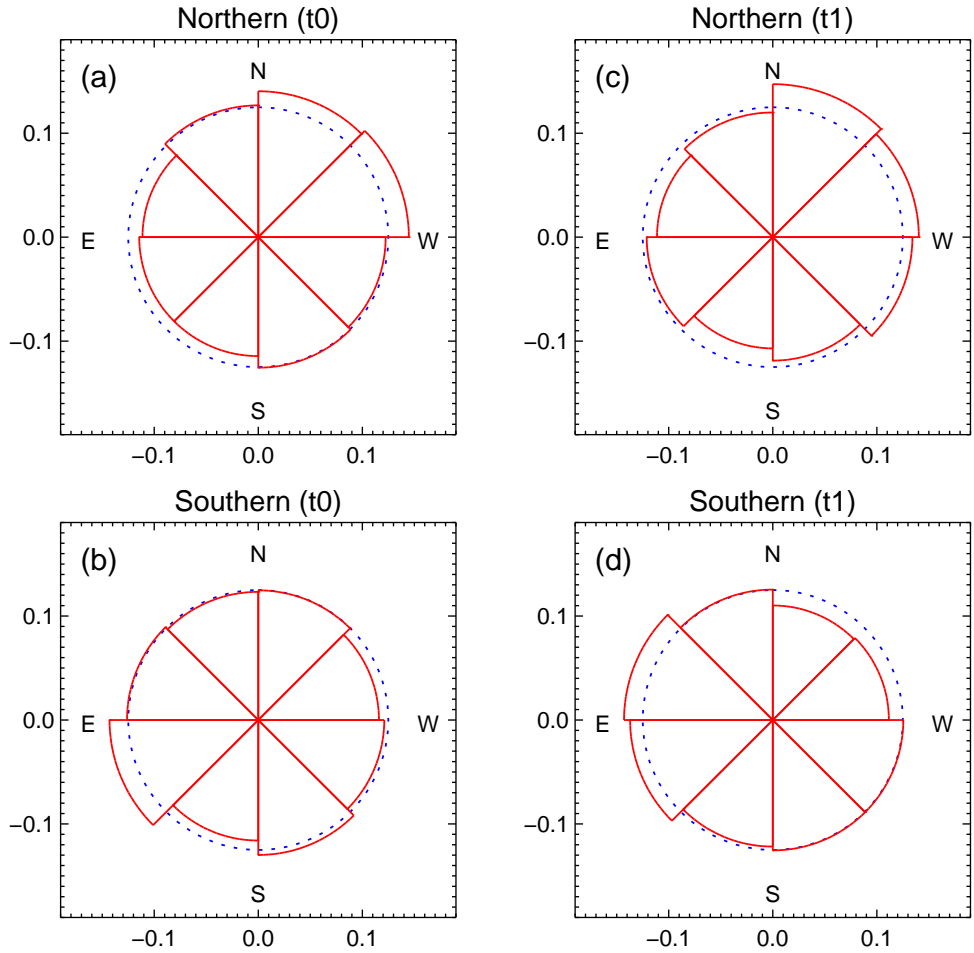}
\caption{Histograms of the ER orientations at the start time (t0;
left column) and the end time (t1; right column) of flux emergence
in angular representation. The northern (upper panels) and the
southern (lower panels) hemispheres are displayed separately. The
dotted blue circles are shown as a reference if the ERs are
definitely randomly oriented. \label{fig10}}
\end{figure}
\clearpage

\begin{figure}
\centering
\includegraphics
[bb=71 207 493 611,clip,angle=0,scale=1.] {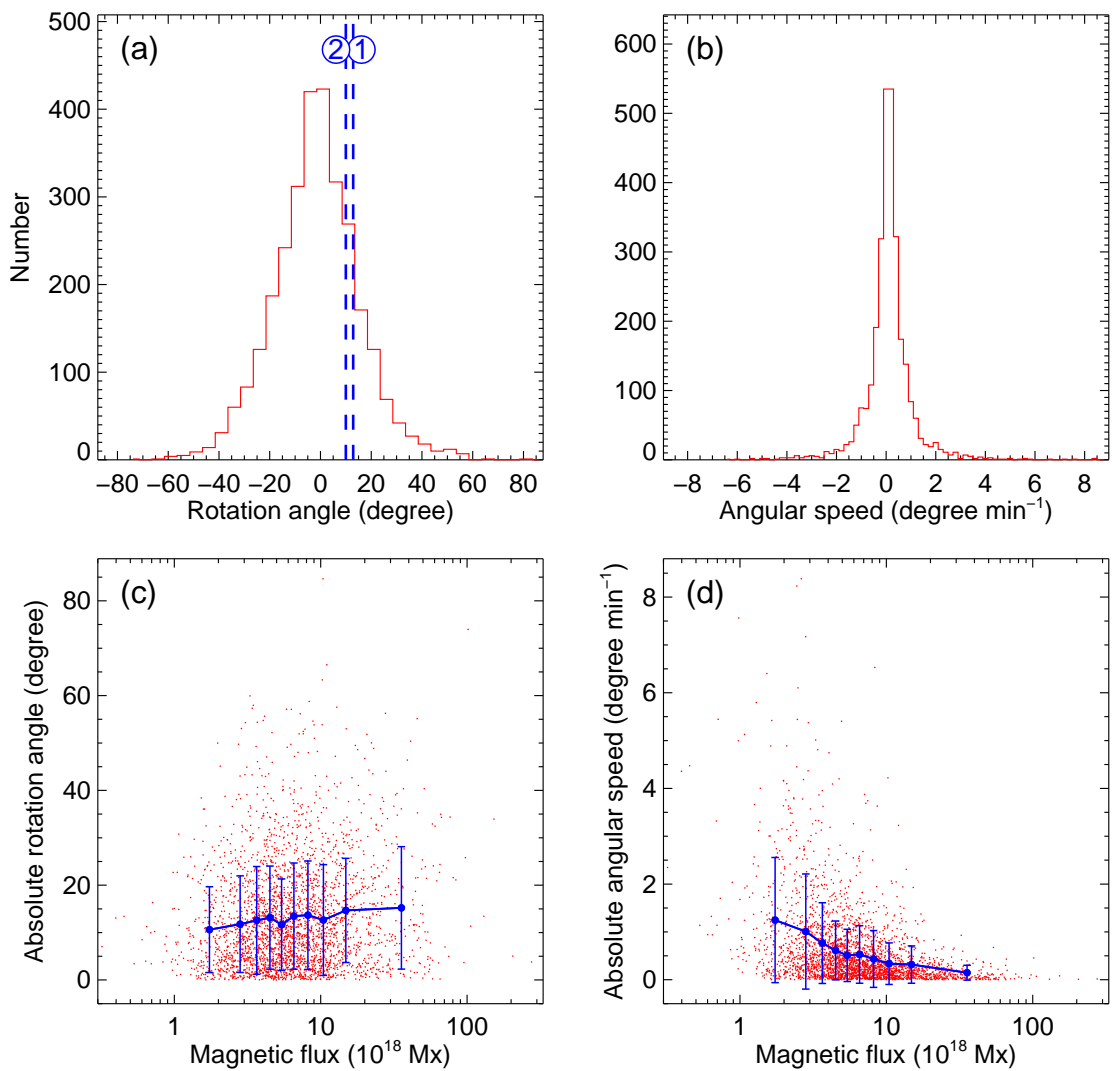}
\caption{\emph{Upper panels}: PDFs of rotation angle (panel (a)) and
angular speed (panel (b)). \emph{Lower panels}: Scatter plots of
absolute rotation angle (panel (c)) and absolute angular speed
(panel (d)) versus magnetic flux of ERs.
Different values obtained by the present study and Wang et al. (2012)
are marked with dashed lines ``1" and ``2", respectively. \label{fig11}}
\end{figure}
\clearpage

\begin{figure}
\centering
\includegraphics
[bb=124 287 449 537,clip,angle=0,scale=1.1] {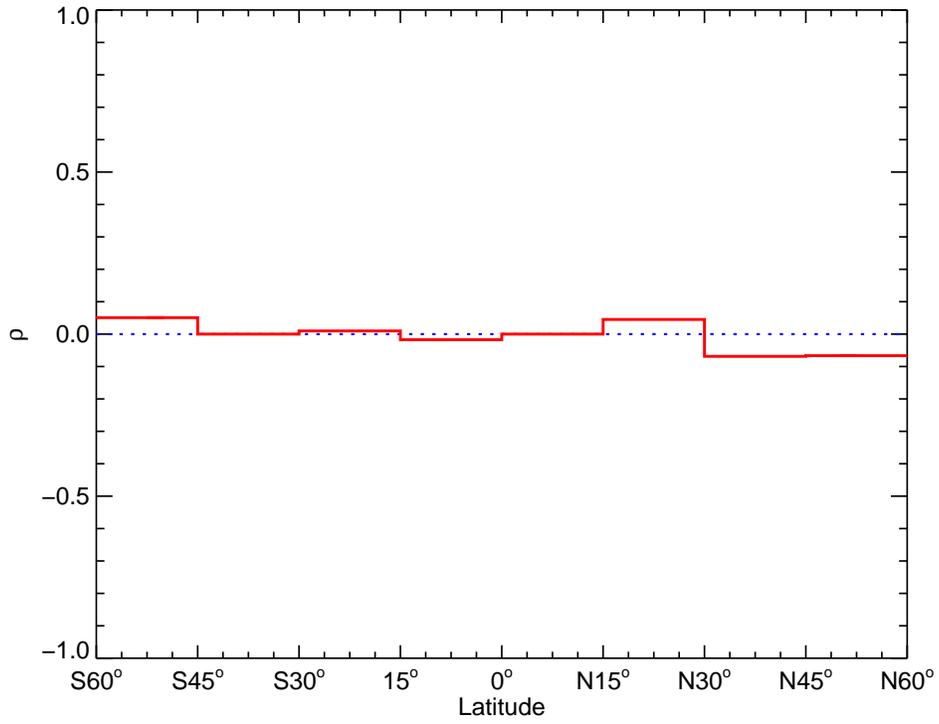}
\caption{Number imbalance between anti-clockwise rotated and
clockwise rotated ERs as a function of latitude. \label{fig12}}
\end{figure}
\clearpage

\end{document}